\documentclass[conference]{IEEEtran}

\usepackage{glossaries}                 
\usepackage{amsfonts}                   
\usepackage{pifont}                     
\usepackage{color-edits}                
\usepackage[table]{xcolor}                      
\usepackage{enumitem}                   
\usepackage{subcaption}                 
\usepackage{graphicx}                   
\usepackage{cleveref}                   
\usepackage{makecell}                   
\usepackage{booktabs}                   
\usepackage{multirow}                   
\usepackage{xfrac}                      
\usepackage{float}                      
\usepackage{dblfloatfix}                
\usepackage{cleveref}                   
\usepackage{dblfloatfix}                
\usepackage{eso-pic}                    
\usepackage{etoolbox}                   
\usepackage{moresize}                   
\usepackage{array}						
\usepackage{soul}  						

\crefname{figure}{Fig.}{Fig.}

\setacronymstyle{long-short}
\newacronym{bow}{BoW}{bunch of wires}
\newacronym{ucie}{UCIe}{universal chiplet interconnect express}
\newacronym{d2d}{D2D}{die-to-die}
\newacronym{ici}{ICI}{inter-chiplet interconnect}
\newacronym{nr}{NR}{non-recurring}
\newacronym{phys}{PHYs}{physical layers}
\newacronym{noc}{NoC}{network-on-chip}
\newacronym{c4}{C4}{controlled collapse chip connection}

\let\mathpi\pi
\renewcommand{\pi}[1]{\textbf{\mathpi{#1}}}

\newif\ifnb     
\nbfalse
\nbtrue

\newif\ifcom    
\comtrue

\newif\ifps     
\psfalse

\addauthor[Summary]{ps}{orange}
\addauthor[Patrick]{pi}{teal}

\newcommand{\ps}[1]{\ifps\pscomment{#1}\fi}
\renewcommand{\pi}[1]{\ifcom\picomment{#1}\fi}

\setlist{leftmargin=1em}

\newcommand{\name}{\emph{FoldedHexaTorus}}

\def\BibTeX{{\rm B\kern-.05em{\sc i\kern-.025em b}\kern-.08em
    T\kern-.1667em\lower.7ex\hbox{E}\kern-.125emX}}

\newcommand{\giturl}{https://github.com/spcl/FoldedHexaTorus}

\begin{document}

\title{FoldedHexaTorus: An Inter-Chiplet Interconnect Topology for Chiplet-based Systems using Organic and Glass Substrates}

\ifnb
	\author{
	\IEEEauthorblockN{Patrick Iff}
	\IEEEauthorblockA{
	\textit{ETH Zurich}\\
	Zurich, Switzerland\\
	patrick.iff@inf.ethz.ch}
	\and
	\IEEEauthorblockN{Maciej Besta}
	\IEEEauthorblockA{
	\textit{ETH Zurich}\\
	Zurich, Switzerland\\
	maciej.besta@inf.ethz.ch}
	\and
	\IEEEauthorblockN{Torsten Hoefler}
	\IEEEauthorblockA{
	\textit{ETH Zurich}\\
	Zurich, Switzerland\\
	torsten.hoefler@inf.ethz.ch}
	}
\else
    \author{\vspace{5em}}
\fi

\maketitle

\begin{abstract}
    Chiplet-based systems are rapidly gaining traction in the market.  
Two packaging options for such systems are the established organic substrates and the emerging glass substrates.  
These substrates are used to implement the inter-chiplet interconnect (ICI), which is crucial for overall system performance.  
To guide the development of ICIs, we introduce three design principles for ICI network topologies on organic and glass substrates.  
Based on our design principles, we propose the novel \name~network topology.  
Our evaluation shows that the \name~achieves significantly higher throughput than state-of-the-art topologies while maintaining low latency.

\end{abstract}

\begin{center}
\vspace{-1em}
\textbf{Code: \giturl}
\vspace{-1em}
\end{center}

\begin{figure*}[t]
\centering
\captionsetup{justification=centering}
\includegraphics[width=1.0\textwidth]{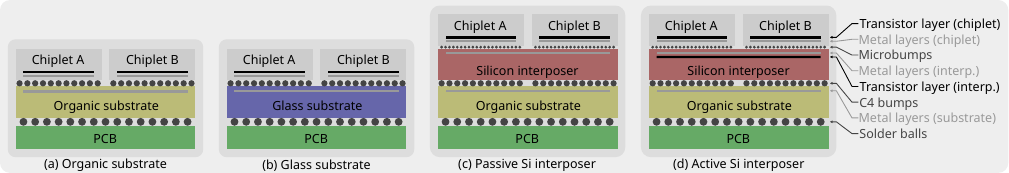}
\caption{\textbf{(\textsection \ref{ssec:back-pack}) Overview of packaging technologies.}}
\label{fig:packaging}
\vspace{-1em}
\end{figure*}

\section{Introduction}
\label{sec:intro}

\ps{Why chiplet-based systems?}

Technology scaling has fueled the ever-increasing performance per cost of processors and accelerators for a long time.  
However, since the $22$\,nm process, each transition to a scaled-down process has been accompanied by a surge in \gls{nr} cost of over $50\%$ \cite{136_design_cost}.  
As a result, designing chips in cutting-edge processes is only economically viable at high production volumes.
Chiplets promise a solution to this problem, as a single chiplet can be reused for multiple products, while the \gls{nr} cost due to design and validation is incurred only once.  
Additional advantages of chiplets include improved yield (and hence lower cost) due to their smaller size compared to monolithic chips, and the option to integrate heterogeneous chiplets (built with different processes) in a single package.

\ps{Challenge of chiplet-based systems and advantages of organic and glass interposers}

Splitting a monolithic chip into multiple chiplets creates the need for a high-throughput \gls{ici}, which is crucial for communication-intensive workloads such as machine learning training and inference or scientific simulations.  
The \gls{ici} is built using \gls{d2d} links \cite{206_d2d_link_1,207_d2d_link_2,286_d2d_link_3}, which are implemented on organic or glass substrates \cite{349_glass_substrate_1,356_glass_substrate_2}, as well as silicon interposers \cite{202_butterdonut} or bridges \cite{229_emib,241_dbhi}.
While silicon interposers and bridges offer higher bandwidth, they come with higher production costs. Therefore, our work focuses on organic and glass substrates.  
Another advantage of these substrates over silicon interposers is that, since they use a different fabrication process, they are not bound by the reticle limit and thus allow the construction of massive systems \cite{348_big_chip}.

\ps{ICI design space for organic and glass interposers}

A major determinant of \gls{ici} throughput is the topology of links between chiplets.  
For systems based on passive silicon interposers or silicon bridges, the \gls{ici} topology is restricted to connecting only adjacent chiplets, resulting in topologies such as \emph{Mesh} and \emph{HexaMesh} \cite{hexamesh}.  
On active silicon interposers, the link length is unrestricted, and many topologies have been proposed \cite{266_double_butterfly, 202_butterdonut, 270_cluscross, 138_kite, 271_sid_mesh}.  
For organic and glass substrates, the link length is less restricted than on passive interposers (due to superior loss characteristics), but more restricted than on active silicon interposers (due to the absence of repeaters), opening up a new and largely unexplored design space for \gls{ici} topologies.

\ps{Our proposal: \name}

In this work, we develop design principles for \gls{ici} topologies on organic and glass substrates (\textbf{contribution 1}).  
These design principles reveal that, to achieve high throughput, the \gls{ici} topology must have a \textit{low network radix}, a \textit{low network diameter}, and \textit{short links}—three properties that are inherently in conflict with one another \cite{359_moore_graphs}.  
By searching for a sweet spot in this design space, we conceive the novel \name~topology (\textbf{contribution 2}).  
The \name~features a constant network radix of six, a constant link length only slightly longer than the chiplet side, and a network diameter of less than $\sqrt{N}$, where $N$ is the total number of chiplets.  
Our evaluation (\textbf{contribution 3}) shows that, for chiplet-based systems with organic and glass substrates, \name~outperforms topologies for passive and active silicon interposers, as well as \gls{noc} topologies.

\section{Background}
\label{sec:back}

\subsection{Overview of Packaging Technologies}
\label{ssec:back-pack}

\textbf{Organic substrates} (see \Cref{fig:packaging}a) are a proven packaging technology with established supply chains.  
They are not bound by the lithographic reticle limit and enable the assembly of systems that surpass the size of monolithic or silicon-interposer-based chips.  
However, the large pitch of \gls{c4} bumps limits the bandwidth of \gls{d2d} links, making them a bottleneck.

\noindent  
\textbf{Glass substrates} (see \Cref{fig:packaging}b) are a packaging technology currently under development \cite{456_blog_glass}.  
Compared to organic substrates, they promise smaller wire and bump pitches, superior thermal stability, and better electrical performance \cite{349_glass_substrate_1, 356_glass_substrate_2}.

\noindent  
\textbf{Passive silicon interposers} (see \Cref{fig:packaging}c) provide higher \gls{d2d} link bandwidth than organic and glass substrates, enabled through the use of fine-pitch microbumps.  
However, passive silicon interposers increase manufacturing costs and complexity, and they suffer from severely limited link length \cite{web_bow}.

\noindent  
\textbf{Active silicon interposers} (see \Cref{fig:packaging}d) alleviate link-length restrictions by providing a transistor layer, enabling the construction of repeaters and buffers \cite{97_intact}.  
However, this transistor layer further increases manufacturing cost and complexity, and can lead to thermal issues.

\subsection{Data Rate and Link Length}
\label{ssec:back-rate-len}

For organic and glass substrates, as well as passive silicon interposers, there is a trade-off between link length and data rate.  
Simulations using the transmission line model \cite{355_transmission_line}, as performed by Kim \cite{346_rate_vs_length}, show that as link length increases, the maximum admissible data rate decreases (see \Cref{fig:rate-vs-length}).  
For passive silicon interposers, the data rate drops significantly when the link length exceeds $4$\,mm.  
For organic and glass substrates, the decline in data rate is less severe and begins only at link lengths of $10$–$20$\,mm.

\section{Design Principles for \gls{ici} Topologies}
\label{sec:dp}

While topologies for data centers \cite{105_slimfly} and \gls{noc}s \cite{44_slimnoc, sparsehamminggraph} are often constructed following graph-theory-based design principles, such principles appear to be lacking for \gls{ici} topologies on organic and glass substrates.
To fill this gap, we introduce three design principles for high-throughput \gls{ici} topologies on organic and glass substrates.

\ps{We want a low diameter}

\vspace{-0.5em}
\subsection{Principle 1: Minimize the Network Diameter}
\label{ssec:dp-diameter}

Minimizing the network diameter (the maximum number of router-to-router hops per packet) has been a major design goal for many topologies in both the data center \cite{357_dragonfly, 105_slimfly} and \gls{noc} \cite{44_slimnoc, sparsehamminggraph} domains.  
The primary motivation for reducing the diameter is that fewer hops per packet translate into fewer packets processed by each router, thereby reducing congestion.  
Recall that in \gls{ici}s based on organic and glass substrates, the limited number of bumps connecting each chiplet to the substrate constitutes a bottleneck.  
While reducing the network diameter is also beneficial for \gls{ici} topologies on silicon interposers, it is even more critical for organic and glass substrates, as it eases pressure on the aforementioned bottleneck and increases throughput.  
Additionally, a smaller diameter results in lower latency, as each chiplet-to-chiplet hop incurs time-consuming processing by \gls{phys} and routers.  
Furthermore, since the energy consumption of a \gls{d2d} link mainly depends on the number of bits transmitted, reducing the network diameter also decreases the overall energy consumption.

\ps{We want short links (1 hop)}

\vspace{-0.5em}
\subsection{Principle 2: Tune the Link-range}
\label{ssec:dp-length}

\textit{Definition}: The \textbf{link-range} is the number of intermediate chiplets that a link stretches across (a link between adjacent chiplets has a range of zero).  
\vspace{0.25em}

\noindent  
As discussed in \Cref{ssec:back-rate-len}, the maximum data rate decreases with increasing link length; hence, shorter links are preferred.  
However, lowering the network diameter (Principle 1) requires a higher link-range, raising the following question:

\begin{center}
\textit{What link-range should be allowed for the best trade-off\\between a low network diameter and short links?}
\end{center}

The answer to this question depends on the packaging technology and the size of the chiplets.  
For passive silicon interposers, the link length is so restricted that only a link-range of zero is practical, leading to the use of topologies such as \textit{Mesh} or \textit{HexaMesh} \cite{hexamesh}.  
For organic and glass substrates, the link length is less restricted, which motivates us to explore the use of higher link-ranges.  
For our analysis, we assume square chiplets of $74$\,mm$^2$ (the same area as in AMD's EPYC and Ryzen processors \cite{92_amd_chiplets}), but we show in \Cref{sec:topo} that our results generalize well when varying the chiplet size.

\begin{figure}[h]
\vspace{-1em}
\centering
\captionsetup{justification=centering}
\includegraphics[width=0.95\columnwidth]{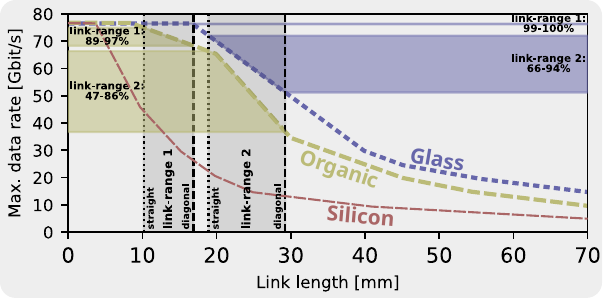}
\caption{\textbf{(\textsection \ref{ssec:dp-length}) Relation between data rate and link length} based on simulations by Kim \cite{346_rate_vs_length}. Yellow and blue areas show the achievable fraction of the max. data rate for a link-range of one and two on organic and glass substrates.}
\label{fig:rate-vs-length}
\end{figure}

\begin{figure*}[b]
\vspace{-1.5em}
\centering
\captionsetup{justification=centering}
\includegraphics[width=1.0\textwidth]{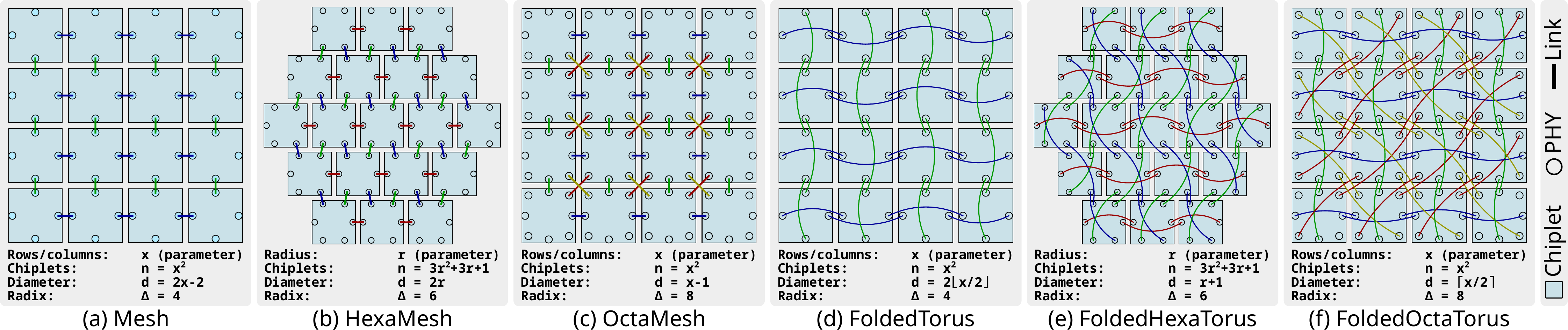}
\vspace{-1.5em}
\caption{\textbf{(\textsection \ref{sec:topo}) Basic \gls{ici} topologies (a-c) and versions optimized for organic or glass substrates (d-f)}.}
\label{fig:topologies}
\end{figure*}

\begin{figure*}[t]
\vspace{-1em}
\centering
\captionsetup{justification=centering}
\includegraphics[width=1.0\textwidth]{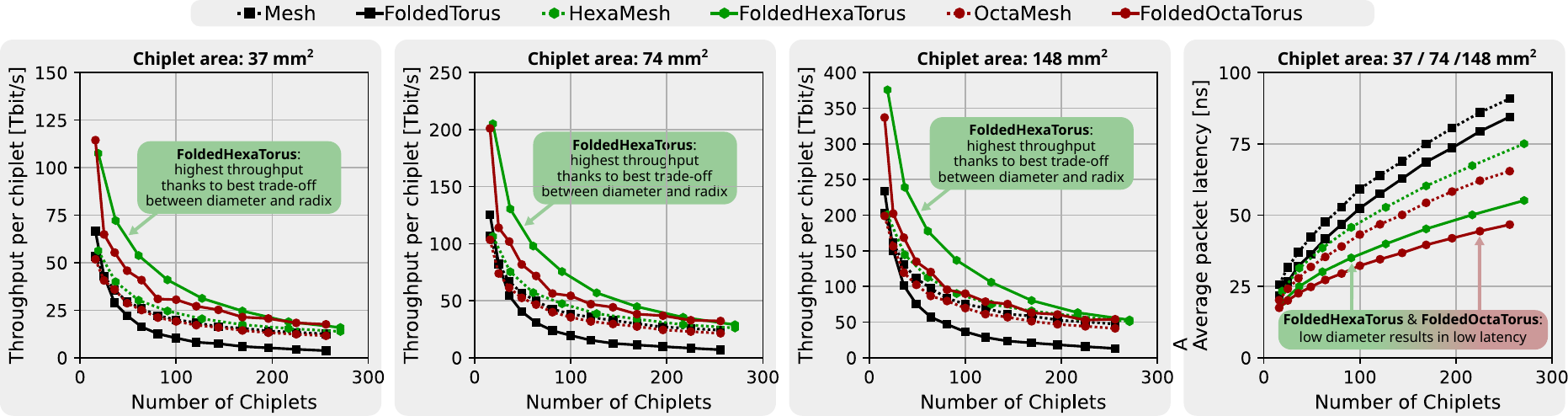}
\caption{\textbf{(\textsection \ref{sec:topo}) Throughput and latency of \gls{ici} topologes.}}
\label{fig:topo-results}
\vspace{-1.5em}
\end{figure*}

\Cref{fig:rate-vs-length} displays the relationship between link length and data rate based on transmission line simulations performed by Kim \cite{346_rate_vs_length}.  
We present a range of possible link lengths (the gray area), spanning from a perfectly straight link (the shortest) to a diagonal link at a $45^\circ$ angle (the longest) for link ranges of one and two.  
For both link-ranges, we show the percentage of the maximum data rate achievable in organic and glass substrates (the yellow and blue areas, respectively).  
We observe that for a link-range of one, links in glass substrates can operate at $99$–$100\%$ of the maximum data rate, while the data rate in organic substrates drops to $89$–$97\%$.  
For a link-range of two, the data rate drops to $66\%$ for glass and $47\%$ for organic substrates in the worst case.  
We conclude that for chiplets of approximately $74$\,mm$^2$, a link-range of one provides the best trade-off between low network diameter and short links.

\subsection{Principle 3: Minimize the Network Radix}
\label{ssec:dp-radix}

The higher the network radix (number of \gls{d2d} links per chiplet), the greater the area overhead of the chiplet, as each \gls{d2d} link requires a dedicated PHY.  
Furthermore, recall that the number of \gls{c4} bumps per chiplet is limited, especially for organic and glass substrates.  
Approximately $50\%$ of these bumps are used for the chiplet's power supply \cite{349_glass_substrate_1}; the remaining bumps are available for \gls{d2d} links and off-chip I/O.  
As the network radix increases, the number of bumps per link—and therefore the per-link bandwidth—decreases.  
Since each link requires a constant number of bumps for non-data wires such as clock or handshake signals ($12$ for \gls{ucie} \cite{web_ucie}), the total off-chiplet bandwidth also decreases with increasing radix.  
These arguments suggest that the network radix should be minimized. However, lowering the network diameter (Principle 1) requires a higher network radix \cite{359_moore_graphs}, creating a conflict between Principles 1 and 3 and raising the following question:

\begin{center}
\textit{What is the optimal balance between minimizing\\network diameter and network radix?}
\end{center}

\noindent
This question is difficult to answer, as the exact relationship between radix and diameter is still an open research problem.
While the Moore bound \cite{359_moore_graphs} provides a theoretical lower bound on the achievable diameter for a given radix and chiplet count, it cannot be directly applied to our setting:
1) for most combinations of chiplet count and network radix, it is unknown whether the Moore bound can actually be achieved, and  
2) the Moore bound assumes links of arbitrary length, whereas we are constrained to links with range one (Principle 2).  
In \Cref{sec:topo}, we address this question experimentally by evaluating different topologies with varying network radices.

\section{Principled Design of \gls{ici} Topologies}
\label{sec:topo}

We follow our design principles (see \Cref{sec:dp}) to construct \gls{ici} topologies for chiplet-based systems using organic or glass substrates.
Consider a \emph{Mesh} topology (see \Cref{fig:topologies}a), which is a common choice for chiplet-based systems (e.g., Tesla Dojo \cite{360_tesla_dojo}).  
We leverage a link-range of one (Principle 2) to transform the \emph{Mesh} into a \emph{FoldedTorus} \cite{104_foldedtorus} (see \Cref{fig:topologies}d), which significantly lowers the network diameter (Principle 1).  
With a low network radix of four, \emph{FoldedTorus} also complies with Principle 3; however, it remains unclear whether further reducing the network diameter at the cost of a higher network radix would improve performance. 
To answer this question, we apply our design principles to two additional topologies.  
\emph{HexaMesh} \cite{hexamesh} is a radix-6 topology (see \Cref{fig:topologies}b) with a lower network diameter than a \emph{Mesh}.  
Again, we use links with range one (Principle 2) to reduce the network diameter, yielding a novel topology that we call \name~(see \Cref{fig:topologies}e).  
As an example of a radix-8 topology, we consider a mesh-like topology with additional diagonal links (see \Cref{fig:topologies}c), named \emph{OctaMesh}.  
By applying design Principles 1 and 2, we transform it into the \emph{FoldedOctaTorus} (see \Cref{fig:topologies}f).

To find the best trade-off between network diameter and network radix, we evaluate the aforementioned topologies using the same methodology as in the main evaluation (see \Cref{sec:eval}). Recall that our analysis motivating the choice of a link range of one (Principle 2) was based on chiplets with an area of $74$\,mm$^2$.
To assess the robustness of our topologies across varying chiplet sizes, we repeat the simulations for chiplets with areas of $37$\,mm$^2$ (half) and $148$\,mm$^2$ (double).

\Cref{fig:topo-results} shows the throughput and latency for chiplet counts ranging from $16$ to $256$ on organic substrates.  
The results for glass substrates are similar, omitted due to space constraints, and can be found in our open-source repository.
For almost all chiplet sizes and chiplet counts considered, \name~achieves the highest throughput while providing the second-lowest latency.  
While throughput depends on various factors such as network radix, network diameter, link lengths (and consequently the maximum data rate), and others, latency is primarily determined by the network diameter.

\Cref{tab:area,tab:power} present the area and power overhead of each topology relative to a \emph{Mesh} topology on organic substrates (results for glass substrates are similar and omitted due to space constraints, they can be found in our open-source repository).
The area overhead results from additional PHYs and thus depends only on the network radix and the chiplet size.  
In contrast, the power overhead (or, in many cases, power savings) relative to a \emph{Mesh} topology depends on the number of bits transmitted per second over the links.  
As a result, it is influenced by both the network diameter (a smaller diameter means each packet traverses fewer links) and the maximum throughput (a topology that achieves higher throughput transmits more packets per second, thereby consuming more power).  
Thus, the power consumption of topologies optimized for \glspl{ici} on organic and glass substrates reflects a trade-off: power savings due to reduced diameter and power overhead due to increased saturation throughput.

\begin{table}[H]
\setlength{\tabcolsep}{8pt}
\centering
\captionsetup{justification=centering}
\vspace{-0.5em}
\begin{tabular}
{
  >{\raggedright\arraybackslash}p{2.0cm} 
  >{\centering\arraybackslash}p{1.5cm} 
  >{\centering\arraybackslash}p{1.5cm} 
  >{\centering\arraybackslash}p{1.5cm}
}
\hline
&\multicolumn{3}{c}{\textbf{Chiplet area relative to a Mesh topology}}\\
\cline{2-4}
\textbf{Topology} & 
\vspace{-0.5em}\textbf{37mm$^2$} &
\vspace{-0.5em}\textbf{74mm$^2$} &
\vspace{-0.5em}\textbf{148mm$^2$} \\
\midrule
Mesh				& 0.00 $\pm$ 0 \%	& 0.00 $\pm$ 0 \%	& 0.00 $\pm$ 0 \%	\\ 	
FoldedTorus			& 0.00 $\pm$ 0 \%	& 0.00 $\pm$ 0 \%	& 0.00 $\pm$ 0 \%	\\
HexaMesh			& 4.34 $\pm$ 0 \%	& 2.27 $\pm$ 0 \%	& 1.16 $\pm$ 0 \%	\\
FoldedHexaTorus		& 4.34 $\pm$ 0 \%	& 2.27 $\pm$ 0 \%	& 1.16 $\pm$ 0 \%	\\
OctaMesh			& 8.69 $\pm$ 0 \%	& 4.54 $\pm$ 0 \%	& 2.32 $\pm$ 0 \%	\\
FoldedOctaTorus		& 8.69 $\pm$ 0 \%	& 4.54 $\pm$ 0 \%	& 2.32 $\pm$ 0 \%	\\
\hline
\end{tabular}
\caption{\textbf{(\textsection \ref{sec:topo}) Total chiplet area (including PHYs) relative to a \emph{Mesh} topology} (mean over all chiplet counts).}
\label{tab:area}
\vspace{-1.5em}
\end{table}

\begin{table}[H]
\vspace{-0.5em}
\setlength{\tabcolsep}{4pt}
\centering
\captionsetup{justification=centering}
\vspace{1.0em}
\begin{tabular}
{
  >{\raggedright\arraybackslash}p{2.0cm} 
  >{\centering\arraybackslash}p{1.9cm} 
  >{\centering\arraybackslash}p{1.9cm} 
  >{\centering\arraybackslash}p{1.9cm}
}
\hline
&\multicolumn{3}{c}{\textbf{Power consumption relative to a Mesh topology}}\\
\cline{2-4}
\textbf{Topology} & 
\vspace{-0.5em}\textbf{37mm$^2$} &
\vspace{-0.5em}\textbf{74mm$^2$} &
\vspace{-0.5em}\textbf{148mm$^2$} \\
\midrule
Mesh				& ~0.00 $\pm$ 0.00	\%	& ~0.00 $\pm$ 0.00 	\%	& ~0.00 $\pm$ 0.00	\%	\\
FoldedTorus			& -0.81 $\pm$ 0.58 	\%	& -1.67 $\pm$ 0.79 	\%	& -3.40 $\pm$ 1.50 	\%	\\
HexaMesh			& -0.12 $\pm$ 0.06 	\%	& -0.35 $\pm$ 0.36 	\%	& -0.74 $\pm$ 0.74 	\%	\\
FoldedHexaTorus		& ~1.19 $\pm$ 1.96 	\%	& ~1.84 $\pm$ 3.21 	\%	& ~2.35 $\pm$ 4.64 	\%	\\
OctaMesh			& -0.83 $\pm$ 0.73 	\%	& -1.69 $\pm$ 1.40 	\%	& -2.93 $\pm$ 2.06 	\%	\\
FoldedOctaTorus		& ~0.28 $\pm$ 1.73 	\%	& -0.10 $\pm$ 2.11 	\%	& -1.66 $\pm$ 2.37 	\%	\\
\hline
\end{tabular}
\caption{\textbf{(\textsection \ref{sec:topo}) Power at saturation throughput relative to a \emph{Mesh} topology} (mean over all chiplet counts).}
\label{tab:power}
\vspace{-1.0em}
\end{table}

Due to its superior throughput, second-lowest latency, and moderate area and power overhead, we recommend \name~as the topology of choice for chiplet-based systems on organic and glass substrates.

\section{Evaluation}
\label{sec:eval}

We compare the throughput, latency, area, and power consumption of our proposed \name~topology against several baseline topologies across a wide range of system sizes. Our analysis spans organic and glass substrates, various architectures (homogeneous and heterogeneous chiplets), and both synthetic traffic patterns and real-world traces.

\subsection{Baseline Topologies}
\label{ssec:eval-baselines}

\begin{figure*}[t]
\centering
\captionsetup{justification=centering}
\includegraphics[width=1.0\textwidth]{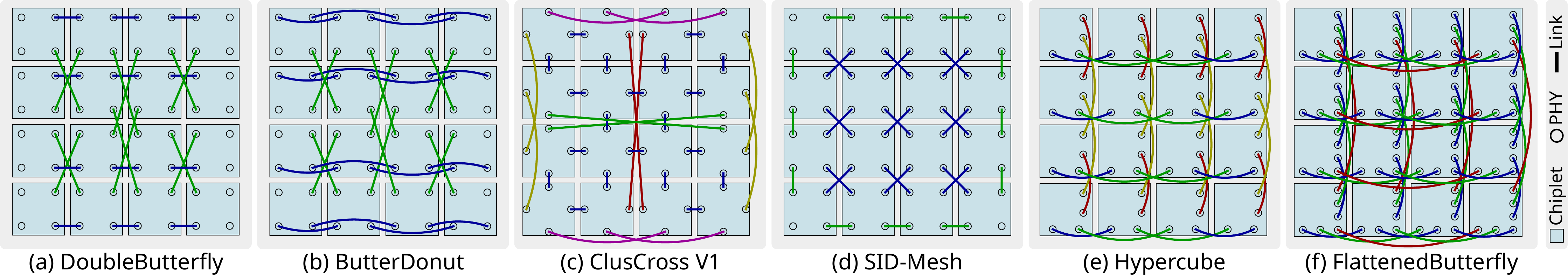}
\vspace{-1.5em}
\caption{\textbf{(\textsection \ref{ssec:eval-baselines}) A selection of baseline topologies; proposed for silicon interposers (a-d) or \gls{noc}s (e-f)}.}
\label{fig:topologies-eval}
\vspace{-1em}
\end{figure*}

\Cref{tab:topologies} lists the baseline topologies we compare against, and \Cref{fig:topologies-eval} visualizes a selection of them.  
While \emph{Mesh} is commonly used in practice and \emph{HexaMesh} has been proposed for both organic substrate- and silicon interposer-based systems, we are not aware of any topologies specifically designed for organic or glass substrates.  
Therefore, our comparison includes a broad spectrum of topologies originally proposed for silicon interposers \cite{266_double_butterfly, 202_butterdonut, 270_cluscross, 138_kite, 271_sid_mesh}, \gls{noc}s in monolithic chips \cite{104_foldedtorus, 106_hypercube, 117_flattenedbutterfly}, and computer networks \cite{403_honeycomb}.  
Since some interposer topologies were originally designed for slightly different architectures, we adapt them to our setting—for example, by using on-chiplet instead of on-interposer routers.
As these topologies are not optimized for organic and glass substrates, most violate at least one design principle: they either feature an unsuitable network radix (Principle 3), an excessive link range (Principle 2), or a suboptimal network diameter (Principle 1).  
\name~is the only topology aligned with all three design principles.

\vspace{-0.6em}
\begin{table}[h]
\setlength{\tabcolsep}{4pt}
\renewcommand{\arraystretch}{1.3}
\centering
\small
\captionsetup{justification=centering}
\begin{tabular}{lccc}
\hline
\textbf{Topology}										& \textbf{Diameter}															& \textbf{Radix}									& \textbf{Link-range}					\vspace{-0.3em}	\\
\midrule
Mesh												& \cellcolor{yellow!15}\footnotesize $2\sqrt{N}-2$							& \cellcolor{green!15}$4$							& \cellcolor{green!15}$0$ 									\\	
Torus 												& \cellcolor{yellow!15}\footnotesize $2\lfloor \sqrt{N}/2 \rfloor $			& \cellcolor{green!15}$4$							& \cellcolor{red!15}\footnotesize $\sqrt{N}-2$ 				\\	
HexaMesh  		\cite{hexamesh}						& \cellcolor{yellow!15}\footnotesize $\frac{\sqrt{12N-3}}{3}-1$				& \cellcolor{green!15}$6$							& \cellcolor{green!15}$0$ 									\\
DoubleButterfly\cite{266_double_butterfly}			& \cellcolor{yellow!15}\footnotesize $\sqrt{N}$								& \cellcolor{green!15}$4$							& \cellcolor{red!15}\footnotesize $\frac{\sqrt{N}}{2}-1$	\\
ButterDonut  	\cite{202_butterdonut}				& \cellcolor{green!15}\footnotesize $\approx \lfloor \frac{2}{3} \sqrt{N} \rfloor	$	& \cellcolor{green!15}$4$					& \cellcolor{red!15}\footnotesize $\frac{\sqrt{N}}{2}-1$	\\
ClusCross V1  	\cite{270_cluscross}				& \cellcolor{yellow!15}\footnotesize $\sqrt{N}-1$							& \cellcolor{green!15}$4$							& \cellcolor{red!15}\footnotesize $\sqrt{N}-2$				\\
ClusCross V2  	\cite{270_cluscross}				& \cellcolor{green!15}\footnotesize $\lceil \frac{3\sqrt{N}}{4} \rceil $	& \cellcolor{green!15}$4$							& \cellcolor{red!15}\footnotesize $\sqrt{N}-2$				\\
Kite Small  	\cite{138_kite}						& \cellcolor{yellow!15}\footnotesize $\sqrt{N}-1$							& \cellcolor{green!15}$4$							& \cellcolor{green!15}$0$									\\
Kite Medium  	\cite{138_kite}						& \cellcolor{yellow!15}\footnotesize $\sqrt{N}$								& \cellcolor{green!15}$4$							& \cellcolor{green!15}$1$									\\
Kite Large  	\cite{138_kite}						& \cellcolor{yellow!15}\footnotesize $\approx\sqrt{N}$						& \cellcolor{green!15}$4$							& \cellcolor{green!15}$1$									\\
SID-Mesh  		\cite{271_sid_mesh}					& \cellcolor{yellow!15}\footnotesize $\sqrt{N}-1$							& \cellcolor{green!15}$4$							& \cellcolor{green!15}$0$									\\
FoldedTrous  	\cite{104_foldedtorus}				& \cellcolor{yellow!15}\footnotesize $2\lfloor \sqrt{N}/2 \rfloor $			& \cellcolor{green!15}$4$							& \cellcolor{green!15}$1$ 									\\	
Hypercube  		\cite{106_hypercube}				& \cellcolor{green!15}\footnotesize $\log_2(N)$								& \cellcolor{green!15}\footnotesize $\log_2(N)$	& \cellcolor{red!15}\footnotesize $\frac{\sqrt{N}}{2} -1 $ 		\\
FlattenedButterfly \cite{117_flattenedbutterfly}	& \cellcolor{green!15}$2$													& \cellcolor{red!15}\footnotesize $2\sqrt{N}-2$		& \cellcolor{red!15}\footnotesize $\sqrt{N}-2$				\\
HoneycombMesh	\cite{403_honeycomb}				& \cellcolor{yellow!15}\footnotesize $1.63\sqrt{N}$							& \cellcolor{green!15}\footnotesize $3$				& \cellcolor{green!15}\footnotesize $0$						\\
HoneycombTorus	\cite{403_honeycomb}				& \cellcolor{green!15}\footnotesize $0.81\sqrt{N}$							& \cellcolor{green!15}\footnotesize $3$				& \cellcolor{red!15}\footnotesize $3\sqrt{N/6}-2$	\\
\midrule
\textbf{FoldedHexaTorus}							& \cellcolor{green!15}\footnotesize $\frac{\sqrt{12N-3}}{6}+\frac{1}{2}$	& \cellcolor{green!15}$6$							& \cellcolor{green!15}1 \vspace{0.3em}						\\
\hline
\end{tabular}
\caption{\textbf{(\textsection \ref{ssec:eval-baselines}) Evaluated topologies}. We highlight diameter, radix, and link-range, to indicate \raisebox{0pt}[0pt][0pt]{\colorbox{green!15}{high}}, \raisebox{0pt}[0pt][0pt]{\colorbox{yellow!15}{moderate}}, or \raisebox{0pt}[0pt][0pt]{\colorbox{red!15}{low}} compliance with design principles.}
\vspace{-2em}
\label{tab:topologies}
\end{table}

\subsection{Evaluation Methodology}
\label{ssec:eval-method}

We measure saturation throughput and average packet latency using the cycle-based BookSim simulator \cite{021_booksim}, which models input-queued, pipelined routers. We use four virtual channels with 4-flit buffers each.
We implement a custom routing algorithm based on Dijkstra's algorithm, incorporating the turn model \cite{turn-model}, the simple cycle breaking algorithm \cite{cycle-breaking}, and a dual graph construction \cite{dual-graph} to enable deadlock-free, shortest-path routing on arbitrary topologies.
\Cref{tab:params} lists the remaining simulation parameters. We extend BookSim to support traffic trace simulation by integrating Netrace \cite{netrace,netrace-tr}. To estimate the area overhead of PHYs and power consumption, we use the RapidChiplet toolchain \cite{rapidchiplet}.

\begin{table}[H]
\setlength{\tabcolsep}{1.5pt}
\centering
\captionsetup{justification=centering}
\vspace{-0.5em}
\begin{tabular}{llllll}
\hline
\textbf{}	& \textbf{Parameter}					& \textbf{Organic} 	& \textbf{Glass}	& \textbf{Reference}							\\
\midrule
$S_c$			& Chiplet spacing						& 150 $\mu$m		& 100 $\mu$m		& \cite{349_glass_substrate_1} (Table 1)	\\
$A_c$			& Chiplet area							& 74 mm$^2$			& 74 mm$^2$			& \cite{92_amd_chiplets} (Page 6)			\\ 
$A_p$			& PHY area								& 0.88 mm$^2$		& 0.88 mm$^2$		& \cite{web_ucie} (Tab. 29)					\\
$P_c$			& Chiplet power							& 25W				& 25W				& Assumption								\\
$E_{bit}$		& Energy per bit						& 0.3 pJ			& 0.3 pJ			& \cite{206_d2d_link_1} (Page 1)			\\
$L_p$			& PHY latency							& 2ns				& 2ns				& \cite{web_ucie} (Table 6)					\\
$L_r$			& Router latency						& 3ns				& 3ns				& Assumption								\\
$f_{pb}$		& Bumps for power						& 50\%				& 50\%				& \cite{349_glass_substrate_1} (Page 3)		\\
$f_{io}$		& Bumps for off-chip I/O				& 20\%				& 20\%				& Assumption								\\
$N_c$			& Cores per chiplet						& 8					& 8					& \cite{92_amd_chiplets} (Page 6)			\\
$P_b$			& Bump pitch							& 50 $\mu$m			& 35 $\mu$m			& \cite{349_glass_substrate_1} (Table 1)	\\
$N_{w}$			& Non-data wires						& 12				& 12				& \cite{web_ucie} (Fig. 73)					\\
$\epsilon_r$	& Dielectric constant 					& 3.1				& 3.3				& \cite{349_glass_substrate_1} (Table 1)	\\
$c$				& Speed of light 						& 299,792 km/s		& 299,792 km/s		& Constant 									\\
\hline
\end{tabular}
\caption{\textbf{(\textsection \ref{sec:topo}) Parameters used in our experiments.}}
\label{tab:params}
\vspace{-1.0em}
\end{table}

\subsubsection{Throughput}
\label{sssec:eval-method-tp}

BookSim reports the relative throughput $T_{r}$, defined as the maximum rate at which each core can inject traffic into the network.  
We compute the absolute per-chiplet throughput $T_{a}$ as follows:
\vspace{-0.25em}
\begin{center}
\includegraphics[width=0.95\columnwidth]{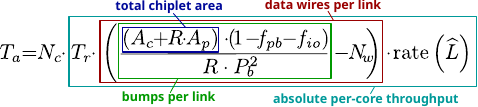}
\end{center}
\vspace{-0.5em}
\noindent
Here, $R$ denotes the router radix, and $\widehat{L}$ represents the maximum link length, computed via RapidChiplet, accounting for both the chiplet spacing $S_c$ and the physical location of a PHY within the chiplet.  
The function $\textit{rate}()$ returns the maximum achievable data rate for a given link length, as defined in \Cref{fig:rate-vs-length}.  
All remaining parameters are listed in \Cref{tab:params}.

\subsubsection{Latency}
\label{sssec:eval-method-lat}

We assume that each chiplet contains a router with latency $L_r$, which can relay messages either to PHYs (with latency $L_p$) or to cores.  
The latency $L_l$ of a link of length $L$ is modeled using the transmission line equation: $L_l = L \cdot \sqrt{\epsilon_r} / c$,  
where $\epsilon_r$ is the relative permittivity of the medium and $c$ is the speed of light in vacuum.  
All parameters are listed in \Cref{tab:params}.  
Since BookSim is cycle-based, we set the cycle time to $1,\text{ns}$ and configure all inputs accordingly.  
The computed link latency $L_l$ is rounded up to the next full cycle.

\subsubsection{Area}
\label{sssec:eval-method-area}

The logic of a chiplet occupies an area of $A_c$, and each PHY contributes an additional area of $A_p$.  
Thus, the total area $A$ of a radix-$R$ chiplet is given by $A = A_c + R \cdot A_p$.

\subsubsection{Power}
\label{sssec:eval-method-power}

The logic of a chiplet consumes $P_c$ watts.  
To estimate the power consumption of each PHY, we count the number of bits transmitted per second during BookSim simulations and multiply it by the energy per bit $E_{bit}$.  
Note that power values are reported at the highest possible throughput.  

\begin{figure}[h]
\centering
\vspace{-0.5em}
\captionsetup{justification=centering}
\includegraphics[width=1.0\columnwidth]{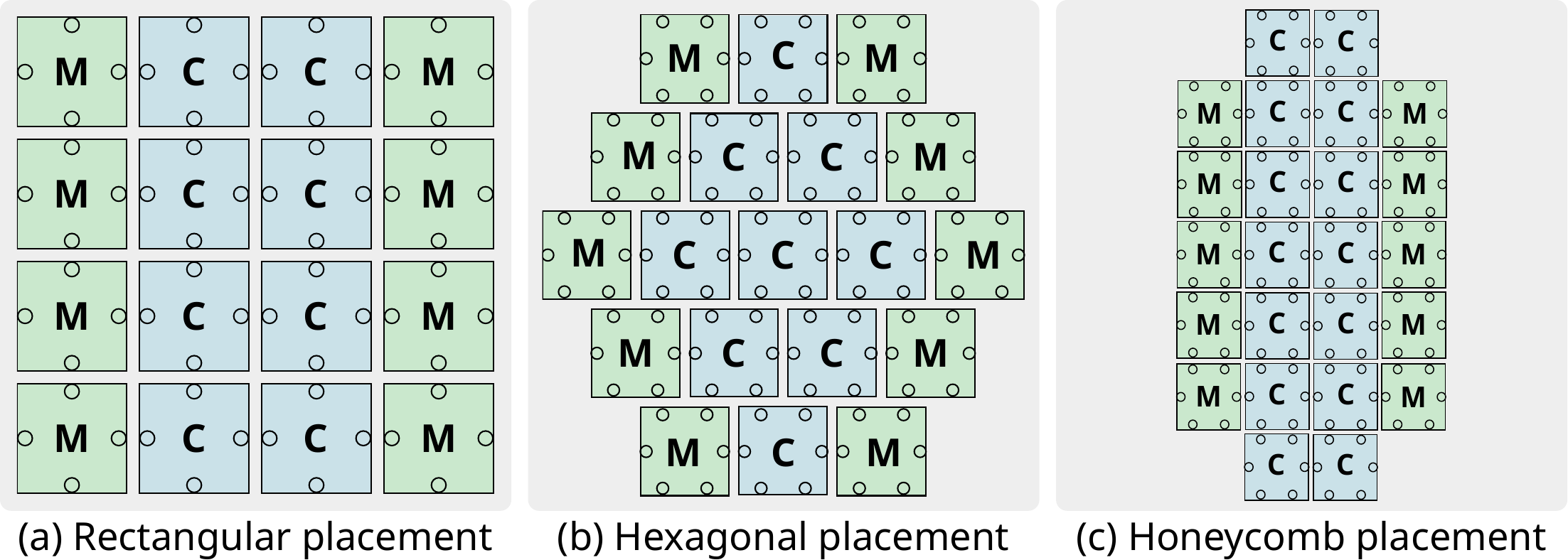}
\vspace{-1.0em}
\caption{\textbf{(\textsection \ref{ssec:eval-uniform}) Placement of compute chiplets (C) and memory chiplets (M) in the heterogeneous architecture.}}
\vspace{-1.0em}
\label{fig:heterogeneous-placements}
\end{figure}
\begin{figure*}[h!]
\centering
\captionsetup{justification=centering}
\includegraphics[width=0.96\textwidth]{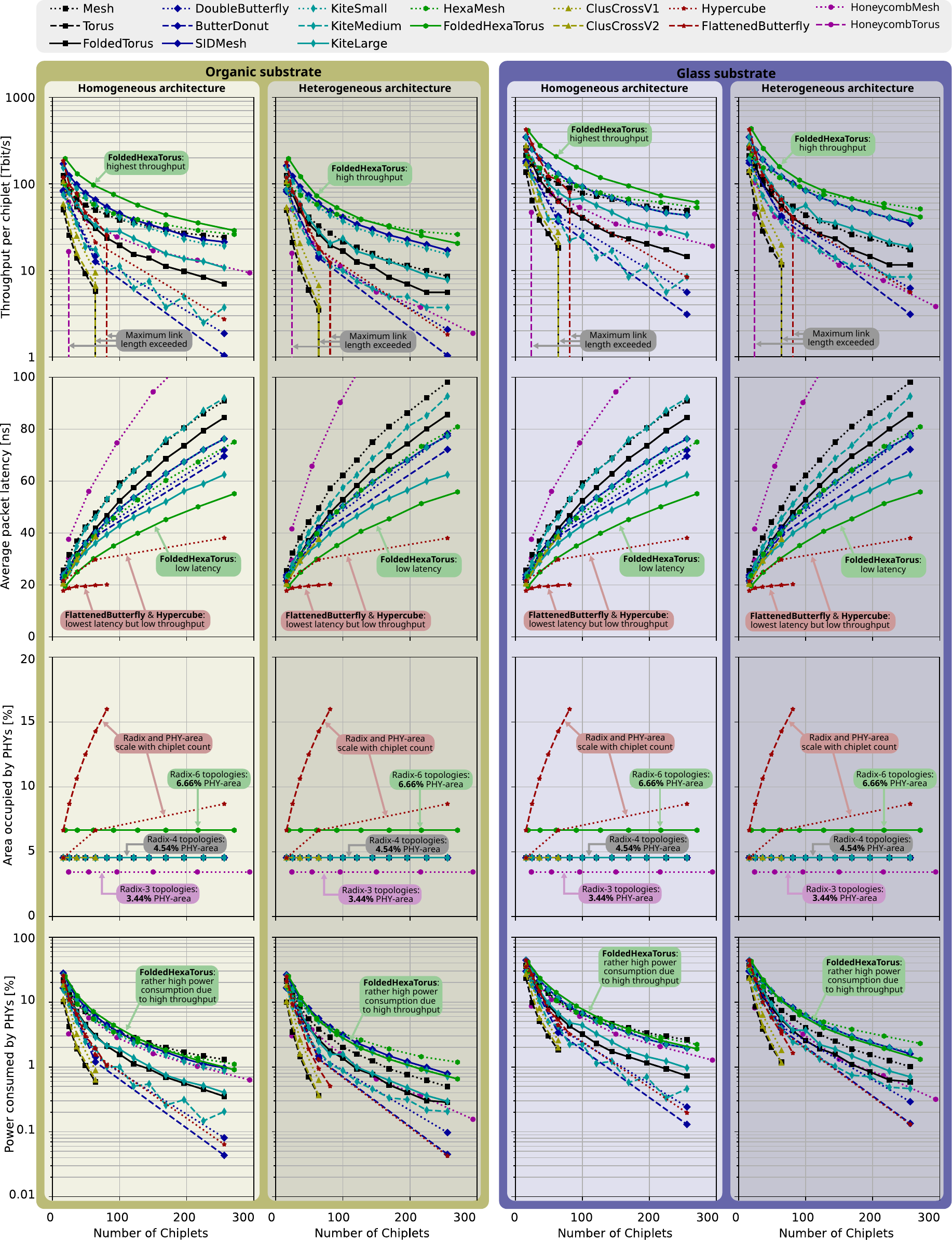}
\caption{\textbf{(\textsection \ref{ssec:eval-uniform}) Throughput, latency, area, and power of \gls{ici} topologies} for varying chiplet counts (random uniform traffic).}
\label{fig:eval-results}
\end{figure*}
\begin{figure*}[h!]
\centering
\captionsetup{justification=centering}
\includegraphics[width=1.0\textwidth]{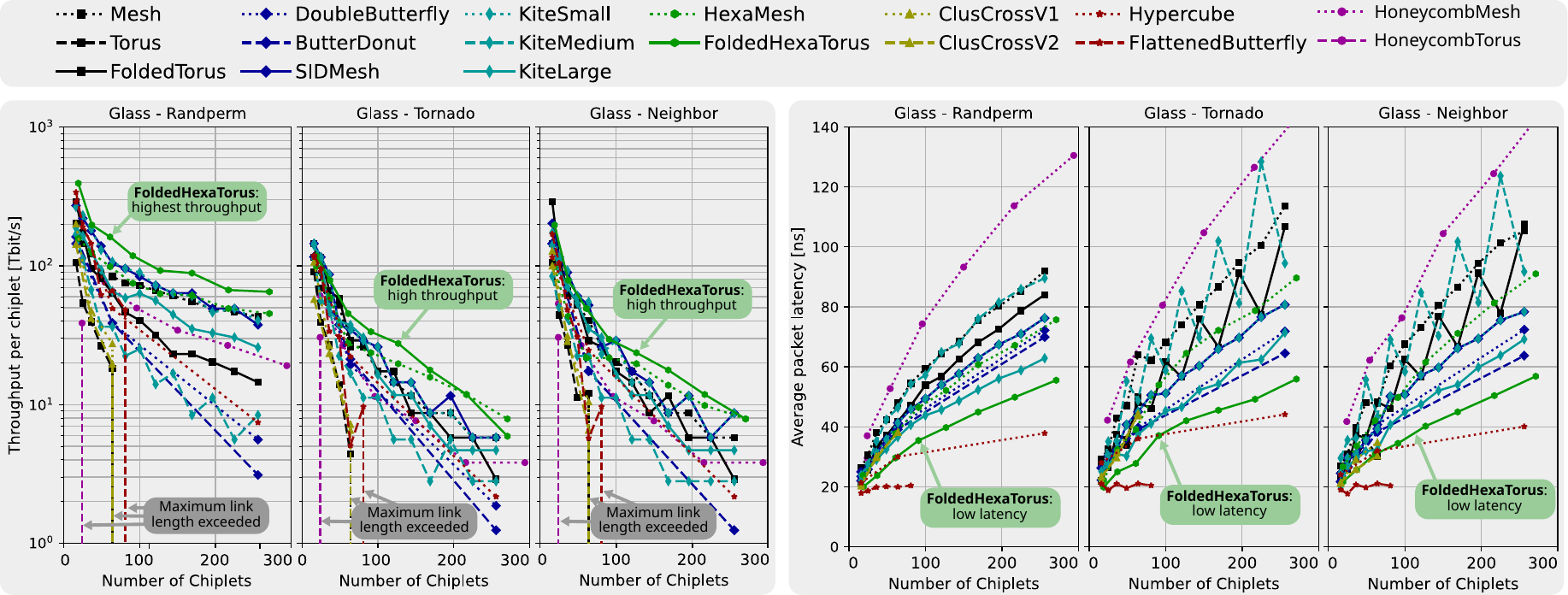}
\caption{\textbf{(\textsection \ref{ssec:eval-synthetic}) Throughput and latency of \gls{ici} topologies for different chiplet counts and traffic patterns.}}
\vspace{-1.5em}
\label{fig:ext-eval-results}
\end{figure*}

\subsection{Results on Synthetic Random Uniform Traffic}
\label{ssec:eval-uniform}

We conduct a broad evaluation using random uniform traffic, as it is as a good proxy for many real-world applications such as graph computations, sparse linear algebra solvers, and adaptive mesh refinement \cite{105_slimfly}.
We evaluate \name~on two architectures: a homogeneous configuration with compute chiplets only, and a heterogeneous configuration with both compute and memory chiplets. In the heterogeneous case, memory chiplets occupy the leftmost and rightmost columns, while the remaining chiplets host compute cores (see \Cref{fig:heterogeneous-placements}).
We use a 50/50 mix of core-to-core and core-to-memory traffic, following common practice \cite{202_butterdonut, 138_kite}.

\Cref{fig:eval-results} compares the throughput, latency, area, and power of \name~against the baseline topologies (see \Cref{ssec:eval-baselines}) for both architectures and for both organic and glass substrates.
We observe that \name~achieves high throughput across all architectures, substrates, and chiplet counts. Notably, for \emph{ClusCross}, \emph{Torus}, \emph{HoneycombTorus}, and \emph{FlattenedButterfly}, throughput drops to zero once the system size exceeds a certain threshold, as some links surpass the maximum permissible length of $70$\,mm.
\name~also demonstrates excellent latency, only outperformed by \emph{FlattenedButterfly} and \emph{Hypercube}, both of which suffer from severely limited throughput.

The percentage of total silicon area occupied by PHYs depends solely on the network radix. Most topologies considered use a radix of four, resulting in a PHY area of $4.54\%$. \name~and \emph{HexaMesh} use a radix of six, slightly increasing the PHY area to $6.66\%$.
The only topologies where the radix scales with the number of chiplets, rather than remaining constant, are \emph{FlattenedButterfly} and \emph{Hypercube}.
Since PHY power consumption is proportional to the number of transmitted bits, it follows the same trend as throughput. As \name~achieves the highest throughput, it also consumes a significant amount of power. However, despite this high throughput, \name's power consumption remains comparable to several topologies with lower throughput.

\subsection{Results on Additional Synthetic Traffic Patterns}
\label{ssec:eval-synthetic}

\Cref{fig:ext-eval-results} shows the throughput and latency of \name~and the baseline topologies under three additional synthetic traffic patterns: \textit{Random Permutation}, \textit{Tornado}, and \textit{Neighbor}. We present results for the homogeneous architecture on a glass substrate; results for the organic substrate are similar and available in our repository.
Performance trends align with those observed under random uniform traffic, with \name~consistently achieving high throughput and low latency across all patterns.

\subsection{Results on Real-World Traffic Traces}
\label{ssec:eval-traces}

\begin{figure}[h]
\centering
\vspace{-0.5em}
\captionsetup{justification=centering}
\includegraphics[width=1.0\columnwidth]{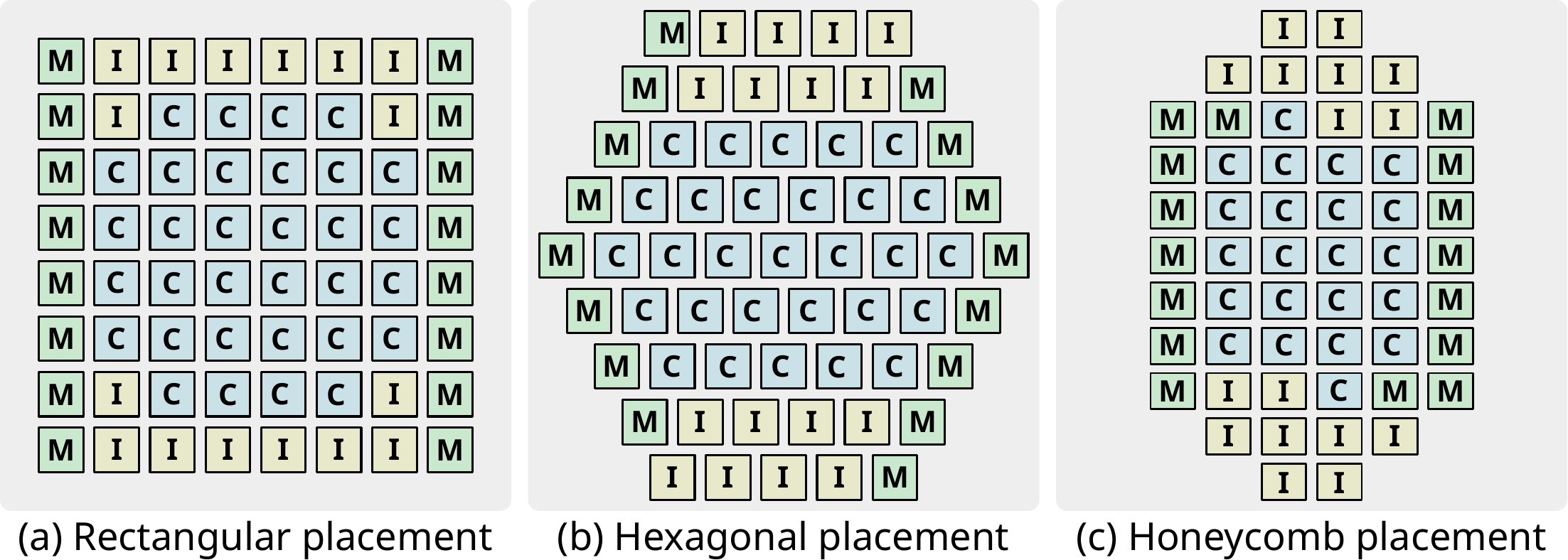}
\vspace{-1.0em}
\caption{\textbf{(\textsection \ref{ssec:eval-traces}) Placement of compute chiplets (C), memory chiplets (M), and IO chiplets (I) used with traces.}}
\vspace{-0.5em}
\label{fig:trace-placements}
\end{figure}

We evaluate \name~using the \emph{blackscholes} and \emph{fluidanimate} traces from the Netrace collection \cite{netrace-traces}, based on the PARSEC benchmark suite \cite{parsec}. Each trace is divided into five regions.
Due to the traces spanning billions of cycles, simulating them in a cycle-accurate simulator is prohibitively time-consuming. Therefore, we simulate the first $100{,}000$ cycles of each region.
All traces include cache coherency traffic between the L1 cache (compute chiplets), L2 cache (memory chiplets), and main memory (IO chiplets). We use an adjusted heterogeneous chiplet placement with compute chiplets in the center, memory chiplets on the left and right, and IO chiplets on the top and bottom (see \Cref{fig:trace-placements}).
Each data packet is split into multiple flits, with the number of flits inversely proportional to the topology's link bandwidth. Control packets are modeled as single flits.

\begin{figure*}[h!]
\centering
\captionsetup{justification=centering}
\includegraphics[width=1.0\textwidth]{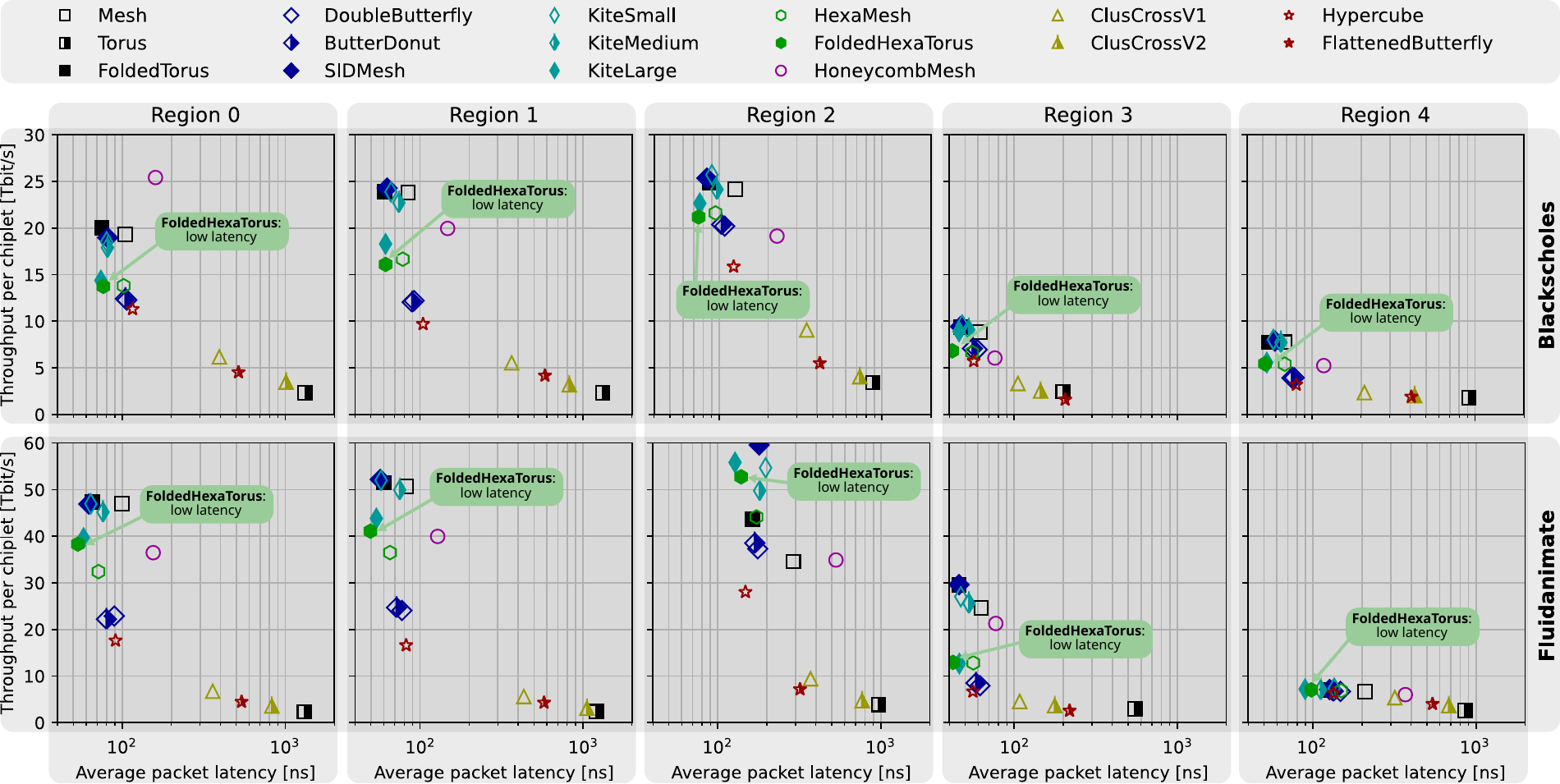}
\caption{\textbf{(\textsection \ref{ssec:eval-traces}) Evaluation on traffic traces}. Points with saturated colors represent Pareto-optimal topologies.}
\vspace{-0.5em}
\label{fig:trace-eval-results}
\end{figure*}

\Cref{fig:trace-eval-results} shows the throughput and latency of \name~and the baseline topologies on an organic substrate for the five regions of the two traces. 
We observe that \name~achieves very low, and in some cases the lowest, packet latency while maintaining reasonable throughput. 

\section{Manufacturing Considerations}
\label{sec:manuf}

While most \gls{ici} topologies assume a rectangular placement of chiplets (see \Cref{fig:heterogeneous-placements}a), we propose using a hexagonal chiplet placement (see \Cref{fig:heterogeneous-placements}b), originally introduced for the \emph{HexaMesh} topology \cite{hexamesh}.  
When applied to systems based on silicon interposers, this hexagonal arrangement results in a mismatch between the rectangular shape of the interposer and the hexagonal layout of the chiplets, leading to a non-negligible interposer area overhead. 
Although manufacturing hexagonal interposers is technically feasible using plasma dicing \cite{096_plasma_dicing} or stealth dicing \cite{095_stealth_dicing}, these methods are less widely adopted than conventional blade dicing, which only supports rectangular shapes.
In contrast, organic substrates are diced using mechanical routing or laser cutting, both of which support arbitrary shapes.  
For glass substrates, mechanical scribbling and breaking or laser cutting are used, with the latter also supporting arbitrary geometries.  
Moreover, since hexagons tessellate the plane without gaps, there is no substrate material waste when producing hexagonal substrates.
In conclusion, while the hexagonal chiplet placement may pose challenges for silicon interposer-based systems, it presents no disadvantages for systems built on organic or glass substrates.

\section{Related Work}
\label{sec:rl}

Chiplet-based systems using organic substrates are an established technology.  
Since AMD \cite{92_amd_chiplets} demonstrated the cost-efficiency of such systems with their 9-chiplet EPYC processors, the number of chiplets per system has grown significantly, reaching 25 in Tesla’s Dojo architecture \cite{360_tesla_dojo}.

Glass substrates, on the other hand, are a more recent technology currently under active development.  
The new opportunities enabled by glass substrates and their associated challenges are well summarized in the work by Usman et al. \cite{356_glass_substrate_2}.  
While Vanna-Iampikul et al. \cite{349_glass_substrate_1} highlight the advantages of glass interposers in terms of area efficiency, wire length, signal integrity, and thermal stability, Kim \cite{346_rate_vs_length} addresses one of their key challenges: power/ground noise.

Regardless of the packaging technology, chiplet-based systems require a high-throughput \gls{ici} to provide sufficient communication bandwidth between chiplets.  
Most prior work has focused on active silicon interposers.  
Jerger et al. \cite{266_double_butterfly} proposed leveraging the interposer’s metal layers to implement an \gls{ici} that handles most of the core-to-memory traffic, using the \emph{DoubleButterfly} topology.  
Kannan et al. \cite{202_butterdonut} later introduced the \emph{ButterDonut} topology, disintegrating the compute chiplet into multiple smaller ones.  
Further developments include the \emph{ClusCross} \cite{270_cluscross}, \emph{Kite} \cite{138_kite}, and \emph{SID-Mesh} \cite{271_sid_mesh} topologies.

One of the few works focusing on passive silicon interposers and organic substrates, rather than active interposers, is \emph{HexaMesh} \cite{hexamesh}, which arranges chiplets in a hexagonal layout and connects each chiplet to its six neighbors.  
With \name, we propose— to the best of our knowledge—the first \gls{ici} topology optimized for organic and glass substrates.

\section{Conclusion}
\label{sec:conc}

Based on our analysis of how network diameter, link-range, and network radix affect an \gls{ici}'s throughput, latency, area, and power, we define three design principles for \gls{ici} topologies on organic and glass substrates:  
1) minimize network diameter,  
2) use a link-range of one, and  
3) minimize network radix.  
Guided by these principles, we propose the novel \textbf{\emph{FoldedHexaTorus}} topology, which has a link-range of one, a network radix of six, and a network diameter below  $\sqrt{N}$, where $N$ is the number of chiplets.  
We evaluate \emph{FoldedHexaTorus} against a broad set of baseline topologies.  
Across system sizes from $16$ to $256$ chiplets and for both organic and glass substrates, it consistently delivers high throughput and near-optimal latency, while incurring only minor area and power overheads.

\ifnb
\section*{Acknowledgements}
\label{sec:ack}

This work was supported by the ETH Future Computing Laboratory (EFCL), financed by a donation from Huawei Technologies.
It also received funding from the European Research Council
\raisebox{-0.25em}{\includegraphics[height=1em]{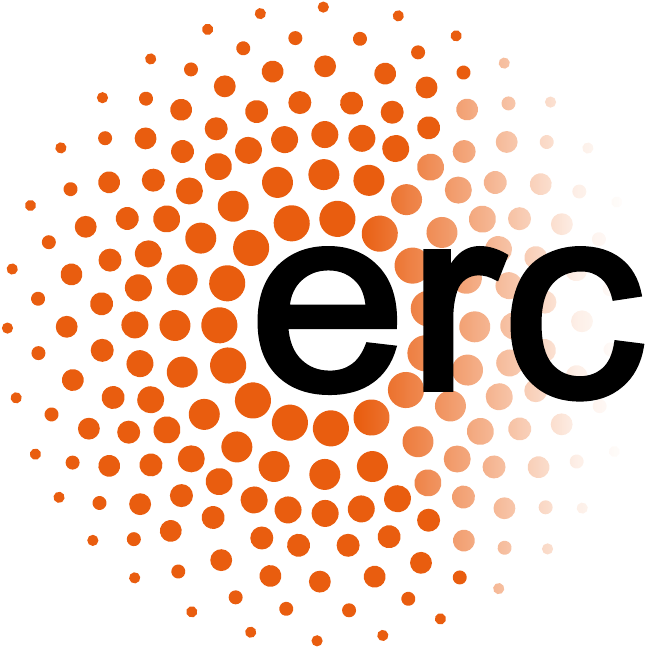}} (Project PSAP,
No.~101002047) and from the European Union's HE research 
and innovation programme under the grant agreement No.~101070141 (Project GLACIATION).

\fi

\bibliographystyle{IEEEtran}
\bibliography{bibliography}

\end{document}